**Observation of coherent oscillation of a single nuclear spin and realization of a two-qubit conditional quantum gate**

F. Jelezko, T. Gaebel, I. Popa, M. Domhan, A. Gruber, J. Wrachtrup

University of Stuttgart, 3. Physical Institute, Stuttgart, Germany

**Abstract**

Rabi nutations of a single nuclear spin in a solid have been observed. The experiments were carried out on a single electron and a single $^{13}$C nuclear spin of a single nitrogen vacancy defect center in diamond. The system was used for implementation of quantum logical NOT and a conditional two-qubit gate (CROT). Density matrix tomography of the CROT gate shows that the gate fidelity achieved in our experiments is up to 0.9, good enough to be used in quantum algorithms.

PACS numbers:

Quantum computers promise to increase substantially the efficiency of solving certain computationally demanding problems like searching databases and factoring large integers. One of the greatest challenges now is to implement basic quantum computational elements in a physical system and to demonstrate that they can be reliably controlled. Quantum gates have been experimentally demonstrated for photons [1], single trapped ions [2,3] and solid state systems like single quantum dots [4] and superconducting charge qubits [5]. Single spins in semiconductors [6], in particular associated with defect centers [7], are promising candidates for practical and scalable implementation of quantum computing even at room temperature [7,8,9]. Such an implementation may also use the reliable and well known gate constructions from bulk nuclear magnetic resonance (NMR) quantum computing [10,11]. Recently for example preparation and detection of entanglement between an electron spin and a nuclear spin in solid have been demonstrated using bulk ESR[12]. It has been shown that a state of a single nitrogen-vacancy color center in diamond can be read out optically [13]. This defect consists of a substitutional nitrogen impurity next to a vacancy in the diamond lattice (see Fig. 1). The defect has been characterized extensively [14] and it was shown that single defect centers can be detected by their strong fluorescence [15,16,17]. The electronic ground and first excited state, are electron spin triplet states (S=1). Optical excitation is only effective between the $m_S=0$ sublevels in both states (see Fig. 1). At low temperature, the spin relaxation time $T_1$ is on the order of seconds and thus a single electron spin state can be detected [13]. In those experiments the fidelity of the state readout is mostly limited by errors associated with photon shot noise and dark counts of the detector. The probability to determine the correct spin state within $T_1$ is around 80%, similar to the case of single ions in traps [18].

If an electron spin is interacting with a paramagnetic nuclear spin, the spin Hamiltonian describing the coupled system is

$$\hat{H} = g_e \beta_e \hat{S}\hat{B} + \hat{S}\vec{D}\hat{S} + \hat{S}\vec{A}\hat{I} - g_n \beta_n \hat{I}\hat{B} \ .$$

Here $\vec{D}$ is the fine structure tensor owing to the interaction of the two uncoupled electron spins, and $\vec{A}$ is the hyperfine interaction tensor related to coupling between the electron and the nuclear spin. The hyperfine coupling of a $^{13}$C (I=1/2) nucleus in the first coordination shell (see Fig. 1) around the defect center is known to be 130 MHz [19]. The natural abundance of $^{13}$C in the samples used is 1.1%. Hence, one out of 30 defect centers should have a $^{13}$C in either of the positions 1 to 3 (see Fig. 1). For demonstration of Rabi oscillations of single nuclear spin and a two qubit conditional quantum gate we have chosen a defect center where the single electron spin is hyperfine coupled to a single $^{13}$C nucleus in one of the nearest neighbor positions 1,2 or 3. A scheme describing the spin levels relevant in this situation is shown in Fig.2. Among the spin levels, 4 transitions are allowed in first order. A and B are electron spin resonance transitions ($\Delta m_S=\pm1$, $\Delta m_I=0$) and C and D are nuclear magnetic resonance transitions ($\Delta m_I=\pm1$, $\Delta m_S=0$). The splitting between states 1 and 2 is determined by the hyperfine coupling of the $^{13}$C nucleus ($\approx130$MHz), whereas the splitting between 3 and 4 is given by the nuclear Zeeman interaction (2-10MHz). Since our detection scheme is only sensitive to the electron spin state, all changes in the nuclear spin states need to be detected via the electron spin. This is equivalent to an optically detected electron nuclear magnetic double resonance (ENDOR) experiment [20].

Fig. 3a shows Rabi nutations of the electron and nuclear spins measured by this technique. In the experiment the electron spin is initialized first by a laser pulse (duration 3μs). After initialization (see Fig. 2b), the system is found either in state 3 or 4. If the system is in state 4, a new initialization is started until state 3 is populated. Starting from this state, frequency selective electron spin resonance pulses with center frequency A are used to drive electron spin Rabi nutations between the states ⏐00> and ⏐10>. To obtain smooth curves, roughly $10^5$ experimental cycles have been averaged. From measurements of the spin nutation decay times or two pulse echo decay curves (data not shown), decoherence times $T_2$ of the electron spin have been determined [21]. Values up to 6 μs were found in our sample, depending on the defect center investigated. It should be noted however, that dephasing times up to 60 microseconds have been reported in literature [22] for samples with low nitrogen concentration. Fig. 3 also shows nuclear spin Rabi nutations between levels $\mid 10>$ and $\mid 11>$ together with the pulse sequence used.

A precise measurement of the nuclear dephasing time has been carried out by recording the Hahn echo decay of a single nuclear spin. Because the nuclear echo is recorded via electron spin, the echo pulse sequence is $\pi/2 - \tau_1 - \pi - \tau_2 - \pi/2$. Fig. 3b show the series of Hahn echos recorded for different delay times. No decay is visible on the time scale of 30 μs. Note that spin memory times of as long as 100 μs have been reported for $^{13}$C nuclei in high purity diamonds [23]. Hence our results show that strong coupling to the electron spin does not induce decoherence of single $^{13}$C nuclear spin and those spins may be of use for solid state quantum computing [24].

The observation of Rabi nutations on transitions A and C provides the basis for a conditional 2 qubit quantum gate. For this gate one qubit is inverted depending on the state of the other qubit. Here, we realized a CROT gate [25], which is equivalent to a CNOT gate, except for a $\pi/2$ rotation of the nuclear spin around the z-axis [26]. However, unitary transformations like the CROT gate are easier to perform than the CNOT gate [27]. In our experiments we have chosen the electron spin as control bit and the nuclear spin as the target bit. The CROT gate is then realized by a $\pi$-pulse on transition C. The results of the CROT gate is state $\mid 11>$ if the qubit has been $\mid 10>$, before the application of the gate (see Fig. 2b).

In order to check the quality of the state prepared by the CROT gate in our experiment, density matrix tomography of the state after the gate has been carried out. To this end, a series of measurements on the diagonal as well as off-diagonal elements of the density matrix have been performed. For measurement of the diagonal elements the signal strength of the transitions A to D have been measured and normalized to the respective signal intensities of the initial state. The off-diagonal elements related to single quantum coherences have been reconstructed by first applying a $\pi/2$ pulse on the transition where the coherences should be measured. Subsequently, the amplitude of the Rabi nutations on the respective transition has been used to calculate the off-diagonal elements. One or two quantum coherences were first converted into measurable zero quantum coherences before their values have been determined. Errors associated with decoherence during this conversion procedure have been taken into account. An example of the density matrix reconstruction is shown in Fig. 4. The density matrix tomography shows the state of the system after a $\pi$ pulse on transition A and subsequent application to the CROT gate ($\pi$ pulse on transition C). Tomography and calculation show that density matrix is almost symmetrical and that the imaginary part is very small. We thus only show the real part of the density matrix. In the ideal case, without decoherence and perfect pulse angles, the only non-zero matrix element of the density matrix after the CROT gate should be $\rho_{22} = 1$, provided that in the initial density matrix $\rho_{11} = 1$. However, in the present case the dephasing and finite linewidth need to be considered. This is why Fig. 4 also shows a numerical simulation of the density matrix after the gate. For a realistic comparison between experiment and theory, a simulation of the density matrix in Fig. 4 has been carried out by calculating $\rho(t)=S^{-1}\rho(t=0)S$ [28], where S is a unitary matrix describing the action of the pulses on the spins in the rotating frame. By taking into account the linewidth of transition A and C as well as the measured dephasing time and the pulse length used, the simulation reproduces the experiment well. It should be pointed out that the gate fidelity can be increased by future technical improvement like a smaller coupling loop, which will substantially increase nutation Rabi frequency. The gate quality presented in Table 1 is measured without considering the overall single shot detection efficiency of 0.8 by the fidelity F given by F= Tr[$\rho_P(t)\ \rho_I(t)$] [4,29], where $\rho_P(t)$ is the measured density matrix and $\rho_I(t)$ is the ideal one.

The present results demonstrate the feasibility of single spin solid-state quantum computing technology using defect centers. In the present experiments, two qubits on a single defect have been used, a single electron and a single [13]C nuclear spin. A third qubit, which is present in the system, the [14]N nuclear spin of the nitrogen-vacancy defect, has not been used here. In isotopically enriched diamond, all [13]C nuclei in the first and probably also in the second coordination shell are of potential use for quantum computing, provided that their nuclear magnetic resonance transitions can be separated in frequency. For larger numbers of qubits, different defect centers need to be coupled [8]. This coupling may be achieved via their mutual optical transition dipole moments [30] over distances around 10 nm. It has already been shown that defects can be written into diamond with an electron microscope [31]. Since electron beams can be focused well below one nm, this may provide a technology to fabricate arrays of nm spaced defect centers and hence ensure scalability of the present approach.

We believe that defects in solids are particularly interesting hardware for quantum computing since such system may allow for operation at room temperature. Mainly this is because in diamond and other materials the electronic and nuclear spin dephasing times only weakly depend on temperature. Especially for the nitrogen-vacancy defect a $T_2$ of 60 μs has been reported at room temperature [22]. The duration of the CROT gate is roughly 0.1 μs. Hence even at room temperature up to $10^3$ gate operations are feasible under present experimental conditions. Only single spin state detection under ambient conditions has not been successful up to now. The main limitation is the sensitivity of our setup. Currently, the minimum averaging time to detect the fluorescence change after a microwave $\pi$ pulse is around 3 ms. This is larger than $T_1$ of the electron spin at room temperature, $T_1 \sim 2$ms. An improvement of one order of magnitude in detection efficiency, which may be achieved by more advanced detection methods like $4\pi$ detection and improved index matching, may allow for averaging times less that $T_1$. Hence, the defects may accomplish all basic requirements for quantum computation under ambient conditions.

**Acknowledgments.** The authors thank D. Suter for loan of ESR microresonator, J. Twamley and S. Kilin for helpful discussions. This work was supported by DFG under the framework of project "Quanten-Informationsverarbeitung", Landesstiftung BW, and EU project QIPDDF-ROSES. One of the authors (IP) acknowledges the support of the Graduiertenkolleg Magnetische Resonanz, University of Stuttgart.

**Correspondence.** Correspondence should be addressed to Joerg Wrachtrup (e-mail:j.wrachtrup@physik.uni-stuttgart.de).

Table 1. Fidelity of the CROT gate for various input states.

| Input state | Fidelity |
| --- | --- |
| 1 | 0.89 |
| 2 | 0.89 |
| 3 | 0.88 |
| 4 | 1.0 |

Figure legends.

Figure 1. (a) Atomic structure of the nitrogen-vacancy defect center. The numbers 1,2 and 3 mark those carbon nuclei, which have the largest hyperfine coupling to the electron spin of the defect center. (b) Scheme of electronic and spin energy levels of nitrogen-vacancy center. Arrows indicate spin-selective excitation and fluorescence emission pathways.

Figure 2 Spin energy level scheme and pulse sequence relevant for the experiment. The energy levels (a) describe the interaction of a single electron with a single $^{13}$C nuclear spin in the ground state of the defect. The quantum number of states 3 and 4 are $m_S$=0, $m_I$=+1/2 and –1/2. The states 1 and 2 comprise the two degenerated electron spin states $m_S$=±1 with nuclear spin quantum numbers $m_I$=+1/2 and –1/2. The pulse sequence (b) used in the experiment comprises laser excitation, microwave (MW) and radio frequency (RF) irradiation. The laser is switched off prior to spin manipulation experiment, and is switched on at the end of the spin part to provide optical read-out (dotted line). MW irradiation is in resonance with transition A. MW $\pi$ pulses (shown as dotted line) are used if states 1 or 2 were used as input for the CROT gate. The CROT gate is a radio frequency $\pi$ pulse (shown as solid line) with a center

frequency of 127 MHz. A subsequent selective microwave $\pi$ pulse followed by an optical pulse belongs to the readout.

Figure 3. **A**. Electron and nuclear spin transient nutations of a $^{13}$C coupled nitrogen-vacancy defect. The fit function represents exponentially decaying harmonic oscillations. Deviation from the fit for short ESR pulses is related to pulse imperfections for short MW pulses. The insets show the applied pulse sequences. **B.** Hahn echo of single $^{13}$C nuclear spin measured for different interpulse delays.

Figure 4. Density matrix of the state of the system after the CROT gate. The left part of the figure shows the experimentally determined values, and the right part shows the result of a simulation.

Jelezko et al., Fig. 1 a/b

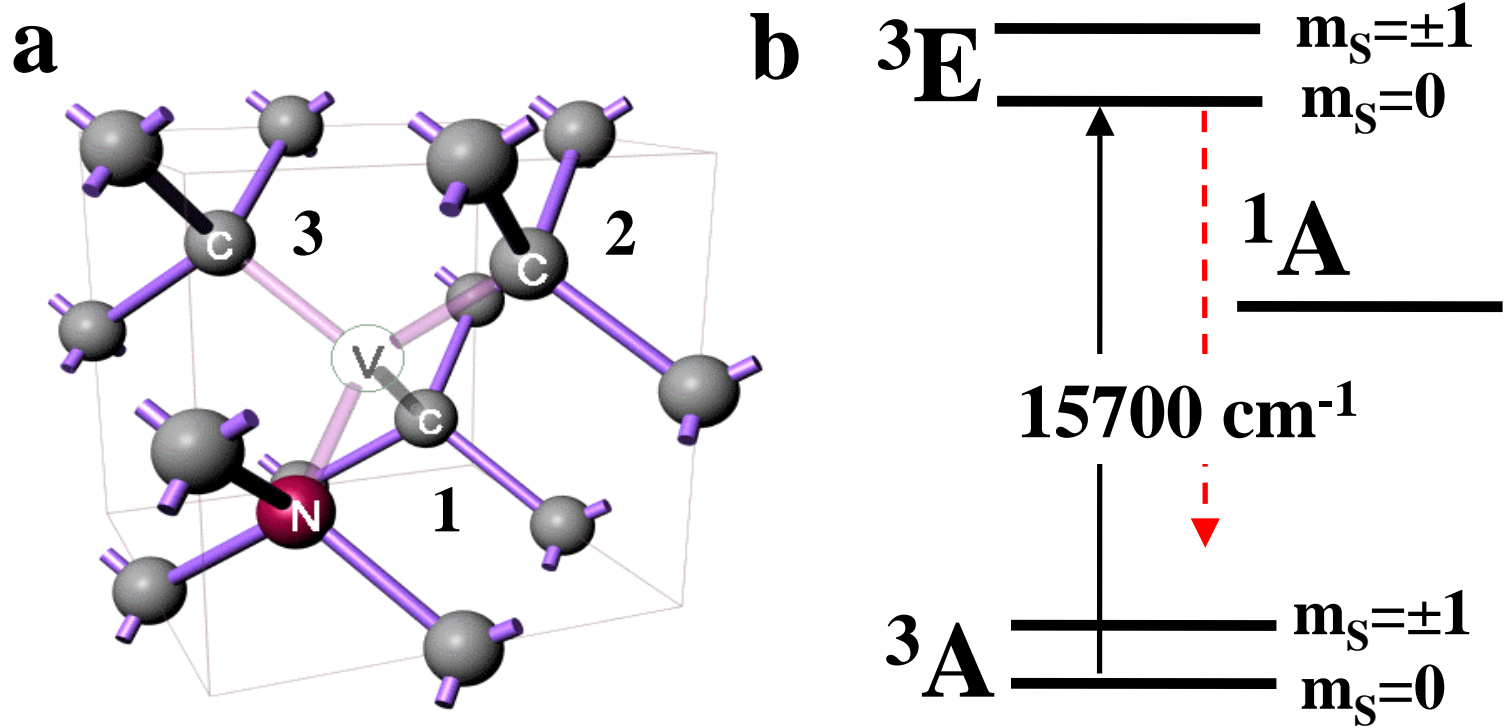

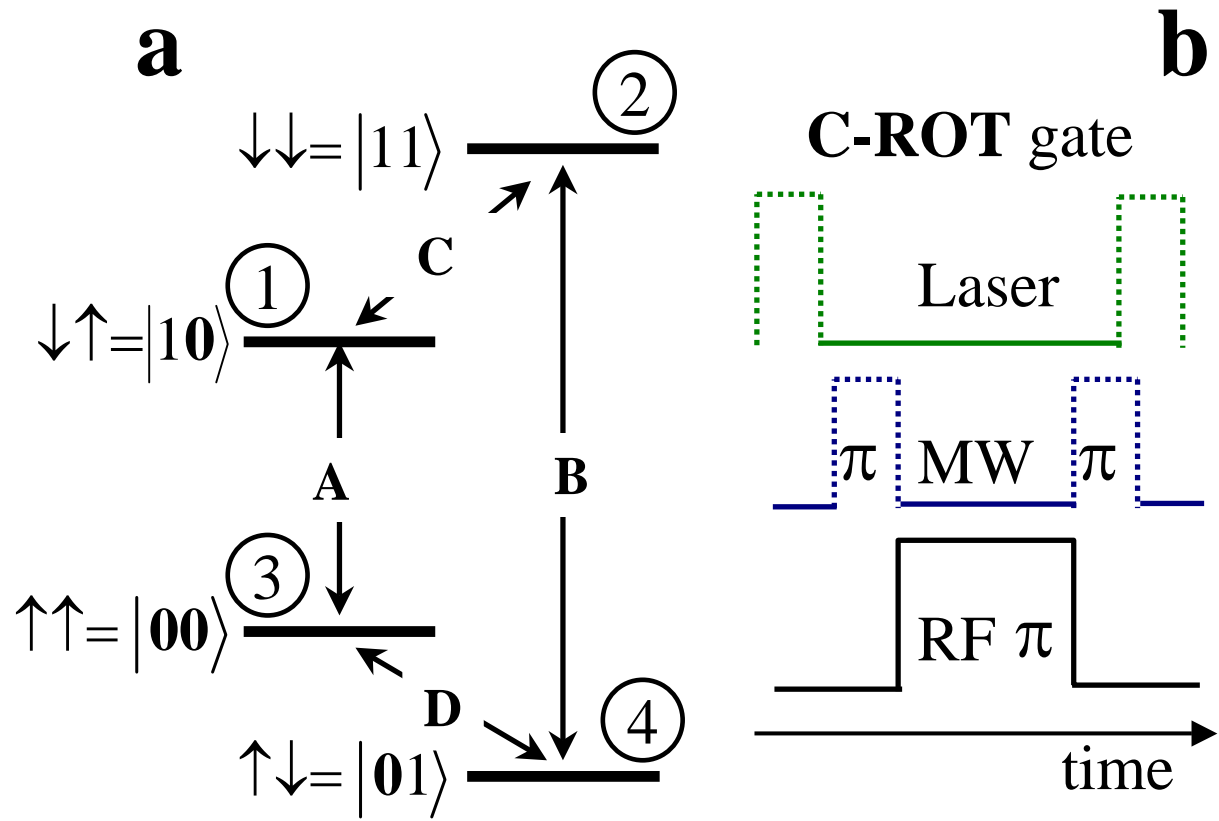

Jelezko et al., Fig. 2 a/b

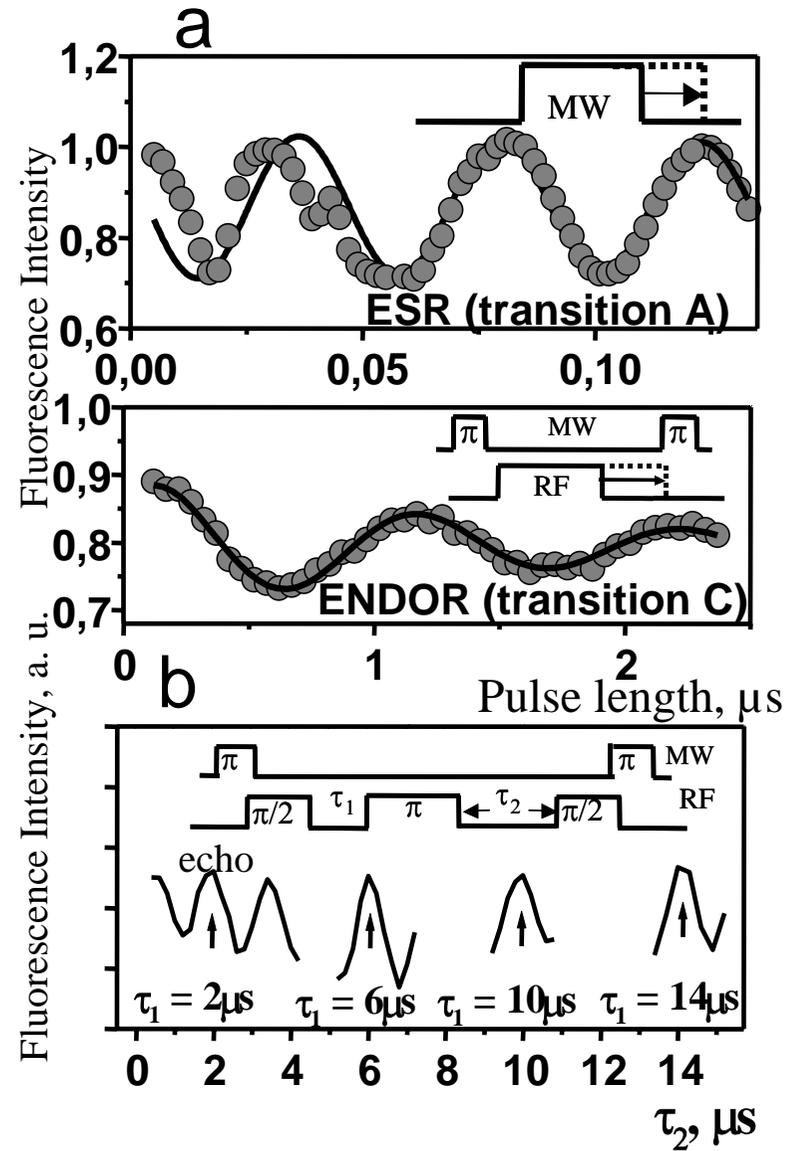

a

Fluorescence Intensity

MW

ESR (transition A)

1,2
1,0
0,8
0,6

0,00        0,05        0,10

π    MW    π

RF

ENDOR (transition C)

1,0
0,9
0,8
0,7

0        1        2

Pulse length, μs

b

Fluorescence Intensity, a. u.

π    MW

π/2    τ₁    π    τ₂    π/2    RF

echo

τ₁ = 2μs    τ₁ = 6μs    τ₁ = 10μs    τ₁ = 14μs

0        2        6        10        14

τ₂, μs

Jelezko et al., Fig. 4

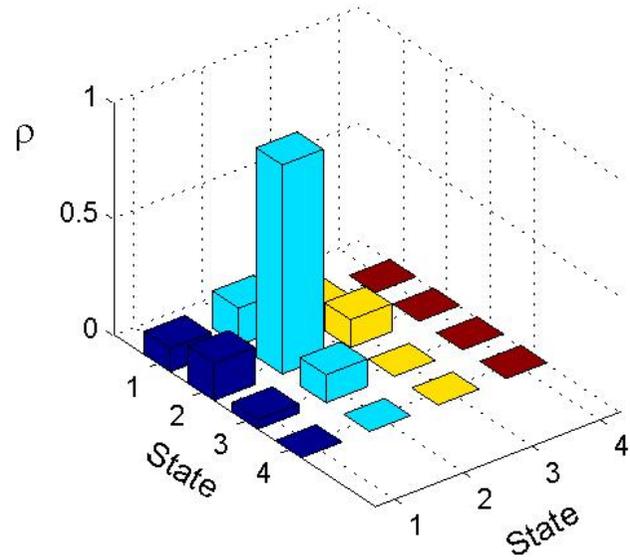
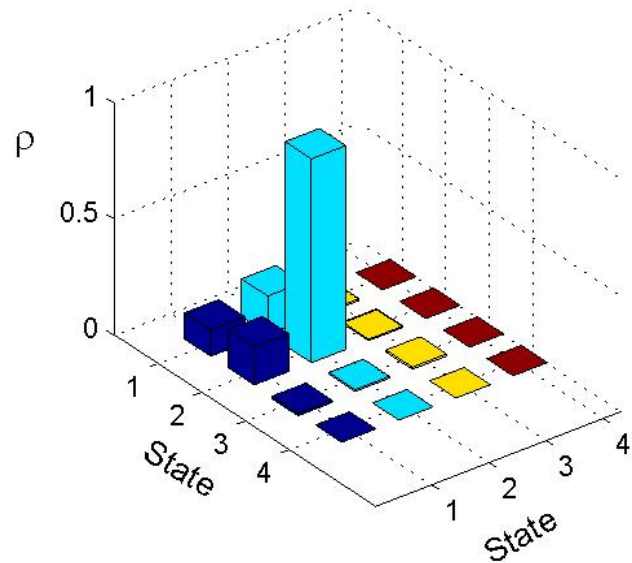